%% file: 2025_IBFD_ADC.tex
\documentclass[conference]{IEEEtran}
\IEEEoverridecommandlockouts

\usepackage{cite}
\usepackage{amsmath,amssymb,amsfonts}
\usepackage{algorithmic}
\usepackage{graphicx}
\usepackage{textcomp}
\usepackage{svg}
\usepackage{psfrag}
\usepackage{comment}
\usepackage{subcaption}
\usepackage{xcolor}
\def\BibTeX{{\rm B\kern-.05em{\sc i\kern-.025em b}\kern-.08em
    T\kern-.1667em\lower.7ex\hbox{E}\kern-.125emX}}

\begin{document}

\title{On Digital Optimization of Analog Self-Interference Cancellation for Full-Duplex Wireless Systems\\[-0.5ex]
\thanks{This work is funded by the German Research Foundation (DFG) under Grant EN 869/4-1 with Project ID 449601577.}
}

\author{\IEEEauthorblockN{
Niklas Knaepper, Gerald Enzner, Aleksej Chinaev}
\IEEEauthorblockA{\textit{Department of Medical Physics and Acoustics, Division of Speech Technology and Hearing Aids,} \\
\textit{Carl von Ossietzky Universität Oldenburg}\\
26111 Oldenburg, Germany \\
\{niklas.knaepper, gerald.enzner, aleksej.chinaev\}@uni-oldenburg.de\vspace{-1ex}}
}

\maketitle

\begin{abstract}
Wireless systems with inband full-duplex transceiver typically require multiple lines of defense against the effect~of harsh self-interference, specifically, to avoid saturation of the analog-to-digital converter (ADC) in the receiver. We may unite the typical tandem operation of successive analog and digital self-interference cancellation (SIC) stages by means of digitally-assisted analog SIC. In this case, the ADC in the receive path requires considerable attention due its possibly overloaded operation outside the intended range. Using neural-network-based architectures of the transmitter nonlinearity, we therefore describe and compare four system options for SIC model optimization with different treatment of the receiver ADC in the learning process. 
We find that omitting the ADC in the backwards path via a so-called straight-through estimation approximation barely impedes model learning, thus providing an efficient alternative to the classical approach of automatic gain control.
\end{abstract}
\vspace*{1ex}


\input{1_intro}
\input{2_prelims.tex}
\input{3_optim_with_ADC}
\input{5_eval.tex}
\input{6_concl.tex}

\bibliographystyle{IEEEtran}
\bibliography{refs_enzner, refs_new}
\end{document}

%% file: 1_intro.tex
\section{Introduction}
\label{sec:intro}

Self-interference cancellation (SIC) is an integral part of inband full-duplex communication, where the same frequency band is used for simultaneous transmitting and receiving. Apart from SIC achieved in propagation domain, e.g., via physical separation of antennas, established SIC components are frequently classified as analog or digital, i.e., using analog filter structures or digital signal processing, respectively \cite{Heino_2015, Herd_2019, Smida2023}. While digital techniques bear more flexibility, higher-order filters and easier implementation, they cannot usually be deployed without help of preceding analog SIC for avoiding saturation of the analog-to-digital converter (ADC) \cite{korpi2014full}.

Digitally-assisted SIC schemes \cite{kiayani2016active, Valkama_2018, liu2017digitally} aim to combine the advantages of digital filtering with those of analog cancellation by modeling the self-interference (SI) channel in digital domain, while realizing the actual cancellation via RF circuitry, i.e., before ADC. Although various publications prove the feasibility of this approach in principle, few explicitly discuss the role of the receiver ADC as part of the signal path during optimization. 
Thus, it is still not obvious how this ADC affects optimization, or, if and how it needs to be accounted for. The latter is especially true at the beginning of optimization, i.e., when still lacking sufficient SIC to avoid heavy saturation.

A straightforward workaround consists in bypassing the low-noise amplifier (LNA) during training \cite{Valkama_2018, kiayani2016active}. However, this may not always be enough to completely avoid saturation, or could even lead to under-utilization of the ADC. 
Although not explicitly addressed as a concern, the authors in \cite{liu2017digitally} seem to avoid saturation during optimization by careful data design. 
Similarly, authors in \cite{kwak2019comparative} choose to reduce the transmission power during training, which, however, potentially changes the nonlinear behavior of the power amplifier. 
The effects of quantization noise and saturation on SIC are discussed in \cite{riihonen2012analog} under the assumption of a known SIC model. Authors in \cite{xing2020comprehensive} provide an elaborate investigation of the effects of ADC resolution on optimization, but in a context of a different digitally-assisted SIC scheme, which does not rely on conventional digital filters. Furthermore,  \cite{riihonen2012analog} and \cite{xing2020comprehensive} mention the possibility to apply an automatic gain control (AGC) as a remedy for ADC saturation, however, missing implementation details, study and discussion in the context of SIC model optimization.

This work investigates the role of the ADC component in digitally-assisted SIC when using (while not critically relying on) neural network models of the nonlinear SI path \cite{enzner2024neural, Stimming2019}. Unfortunately, ADC simulation as a component of a backpropagation model is not immediately a productive approach due to ill-conditioned differentiation properties of its step-function. We hence propose model representation of the ADC within the dynamic range, or even including the saturation range, according to the recent straight-through estimation (STE) theory for quantized networks \cite{bengio2013estimating, yin2019understanding}.
It is demonstrated that STE allows for SIC learning despite heavy initial ADC saturation and it performs similarly to an optimization strategy that additionally relies on AGC. We also compare to a system trained offline on quantized signals with low LNA gain. 
Our experimental comparison includes evaluation of SIC and end-to-end performance in terms of bit error rates (BERs).

The remainder of this paper is organized as follows.~Sec.~\ref{sec:prelims} provides an overview of digitally-assisted SIC. Sec.~\ref{sec:optiWithADC} then describes the approach for system simulation and introduces the different ADC-related learning schemes.
Evaluation results are presented in Sec.~\ref{sec:eval} and Sec.~\ref{sec:concl} concludes.

%% file: 2_prelims.tex
\section{System arrangement and Simulation}
\label{sec:prelims}
\begin{figure}[h]
    \centering
    (a) System overview\\[2ex]
    \psfrag{P}{PA}
    \psfrag{LNA}{LNA}
    \psfrag{DAC}{DAC}
    \psfrag{ADC}{ADC}
    \psfrag{SIC}{SIC}
    \psfrag{Neural Net}{Neural Net}
    \psfrag{r[k]}{$r[k]$}
    \psfrag{e(t)}{$r(t)$}
    \psfrag{s[k]}{$s[k]$}
    \psfrag{ar(t)}{$\alpha r(t)$}
    \psfrag{z(t)}{$z(t)$}
    \psfrag{y(t)}{$y(t)$}
    \psfrag{x(t) Signal of interest}{$x(t)$ Signal of interest (SOI)}
    \psfrag{h(t)}{$h_{\text{SI}}(t)$}
    \includegraphics{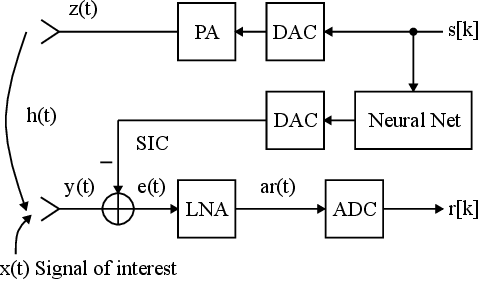}\\[2ex]
    (b) Signal levels\\[2ex]
    \psfrag{n(t)}{$n(t)$}
    \psfrag{y(t)}{$y(t)$}
    \psfrag{z(t)}{$z(t)$}
    \psfrag{passive SIA}{passive SIA}
    \psfrag{PdBm}{P [dBm]}
    \psfrag{-77}{-77}
    \psfrag{-37}{-37}
    \psfrag{-15}{-15}
    \psfrag{0}{0}
    \psfrag{20}{20}
    \includegraphics{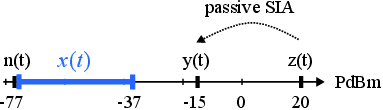}\\
    \caption{IBFD transceiver model with digitally-assisted SIC.}
    \label{fig:modelBlockDiagram}
\end{figure}

A block diagram of the transceiver system, including the digitally-assisted SIC mechanism, is shown in Fig.~\ref{fig:modelBlockDiagram} (a), and Fig.~\ref{fig:modelBlockDiagram} (b) illustrates power levels of the involved signals.
The analog transmission signal $z(t)$ with power $P_z=20$\,dBm results from the digital transmit signal $s[k]$ through digital-to-analog conversion (DAC) and subsequent power amplification (PA) with nonlinearity. Note that this work does not consider issues related to the transmitter DAC, since these are typically not of concern.
The transmit signal is received as
\begin{equation}
y(t) = h_{SI}(t) * z(t) + x(t)    
\end{equation}
over the linear SI channel $h_{SI}(t)$, where $*$ denotes linear convolution, and $x(t)$ is the typically smaller signal of interest (SOI). We assume SOI-SNR in the range of [0, 40]\,dB with respect to a receiver noise floor around
$P_n=-77$\,dBm not explicitly shown in Fig.~\ref{fig:modelBlockDiagram} (a).

A neural network models the nonlinear transmission path from signal $s[k]$ to $y(t)$. It uses $s[k]$ to predict, after auxiliary DAC, the received SI signal $y(t)$, such that SIC can be achieved via subtraction. Our internal model architecture conforms to that of the "global" Hammerstein model described in \cite{enzner2024neural}, i.e., comprising a small multi-layer perceptron (MLP) and a subsequent linear layer. Our network output is scaled to account for the SI signal power of $P_y=-15$\,dBm (due to passive isolation) such that the network internally operates around $0$\,dBm level to ease training conditions.
Following the SIC, the residual $r(t)$ passes through the receiver LNA with gain $\alpha$ and finally the receiver ADC. It becomes available as $r[k]$, the digital residual driving model optimization.

For generating data, we employ the procedure for simulating a time-invariant SI system detailed in \cite{enzner2024neural}. In particular, $s[k]$ simulates orthogonal frequency division multiplexing (OFDM) transmission according to the WLAN High-Throughput (HT) standard IEEE-802.11 \cite{Qureshi_2023} in complex-basedband representation with 20MHz bandwidth, 16-QAM, and coderate 3/4.

Operation of the presented transceiver system, including analog SIC, is simulated in software. To this end, the analog quantities $z(t)$, $h_{\text{SI}}(t)$, $y(t)$ and $r(t)$ are approximated digitally as $z_a[k]$, $h_{\text{a,SI}}[k]$, $y_a[k]$ and $r_a[k]$, respectively, using 64-bit floating-point representations. In contrast, the receiver ADC, i.e., the primary concern in this work, uses a maximum resolution of $B=12$\,bit, thus enabling meaningful simulation of quantization errors. Once the signal has passed the ADC, we return to 64\,bit word length for further processing.

Quantization and saturation of the receiver ADC are applied individually to the I/Q components of the complex baseband signal
using the mid-rise step-function
\begin{equation}
    \text{ADC} \big\{ \alpha\,r_{\text{a,IQ}}[k] \big\} = \Delta \cdot \text{round} \bigg( \frac{g(\alpha\,r_{\text{a,IQ}}[k])+\Lambda}{\Delta}  \bigg) - \Lambda \; ,
    \label{eq:midRiseQuant}
\end{equation}
with $\Lambda$ the absolute saturation threshold, $\Delta=2\Lambda/(N-1)$ the quantization stepsize, $N=2^B$ the number of quantization steps for $B$ bits, and $g(\cdot)$ a clipping function
\begin{equation}
g(x)=\left\{\begin{array}{l}
-\Lambda, ~x\leq-\Lambda \\
x, ~-\Lambda<x<\Lambda,\\
\Lambda, ~x \geq \Lambda.
\end{array}\right.
\label{eq:ADCsat}
\end{equation}
Note that in (\ref{eq:midRiseQuant}), the shift by $\Lambda$ and $-\Lambda$ before and after rounding, respectively, is used to express a mid-rise as opposed to a mid-tread quantization step-function.

%% file: 3_optim_with_ADC.tex
\section{Simulated SIC Model Optimization with ADC}
\label{sec:optiWithADC}

We now turn our attention to the optimization of the neural network part in the context of the entire system in Fig~\ref{fig:modelBlockDiagram}.
Due to our all digital simulation of the system, backpropagation would theoretically be able to traverse the simulated ADC layer with its digital input and output. In order to still account for 
\begin{figure}[h]
    \centering
    \psfrag{ADC}{ADC}
    \psfrag{ar[k]}{$\alpha r_a[k]$}
    \psfrag{r[k]}{$r[k]$}
    \psfrag{dr[k]}{$\Delta r[k]$}    
    \includegraphics{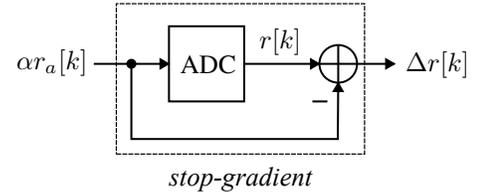}
    \caption{''ADC Noise'' block by ''stop-gradient'' encapsulation.}
    \label{fig:noGradQuantNoiseBlock}
\end{figure}
a realistic scenario, i.e., with the analog ADC input and the ADC layer inaccessible to the backpropagation algorithm, 
we employ \textit{Tensorflow's}~\cite{abadi2016tensorflow} \textit{stop-gradient} (SG) function, as illustrated in Fig.~\ref{fig:noGradQuantNoiseBlock}. 
Specifically, we encapsulate computation of the digitization error as
\begin{equation}
    \Delta r[k] = \text{SG} \big( \text{ADC} \big\{ \alpha r_a[k] \big\} - \alpha r_a[k] \big) \;,
    \label{eq:digError}
\end{equation}
comprising both quantization noise and saturation error. Our digital residual after ADC is then simulated additively as
\begin{equation}
    r[k] = \alpha r_a[k] + \Delta r[k]\;.
\end{equation}

In this way, the ADC layer is deliberately isolated from the backwards pass, whereas its distortion effects are indeed provided in the digital residual $r[k]$, thereby realistically influencing the evaluation of gradients during backpropagation. With this technique as a foundation, we consider different strategies to cope with ADC effects during optimization.

\subsection{Backpropagation through ADC (BPAD)}
\label{sec:bpad}

Looking at Fig.~\ref{fig:modelBlockDiagram}, 
backpropagation for minimization of $r[k]$ would intuitively consider 
the ADC as a nonlinear layer that acts upon the analog residual signal $r(t)$. In practice, this is not possible, since (i) the gradient of the ADC step-function is zero almost everywhere and (ii) the input to the ADC layer, namely $r_a[k]$, is not observable and hence not known. 

\begin{figure}[h]
    \centering
    \psfrag{Neural Net}{Neural Net}
    \psfrag{ADC}{ADC}
    \psfrag{ADC Noise}{ADC Noise}
    \psfrag{LNA}{LNA}
    \psfrag{SIC}{SIC}
    \psfrag{agra}{\!\!$g(\alpha r_a[k])$}
    \psfrag{y[k]}{$y_a[k]$}
    \psfrag{ar[k]}{$\alpha r_a[k]$}
    \psfrag{s[k]}{$s[k]$}
    \psfrag{r[k]}{$r[k]$}
    \psfrag{dr[k]}{$\Delta r[k]$}
    \includegraphics{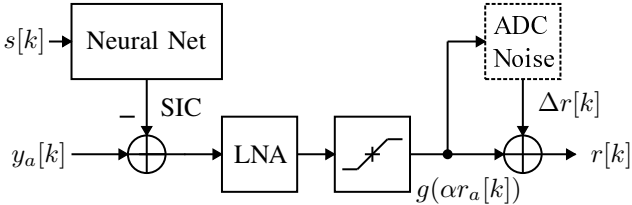}
    \caption{Setup for simulated training with partial ADC model.}
    \label{fig:BPAD}
\end{figure}

We may consider partial ADC simulation to at least account for saturation effects, i.e., according to (\ref{eq:ADCsat}), which leads to the arrangement in Fig.~\ref{fig:BPAD}. The LNA output is passed through a saturation layer $g(\cdot)$, which represents the functional part of the ADC that is within the dynamic range well conditioned for backpropagation. The remaining quantization noise is then added via the previously detailed ''ADC Noise'' block fed with already saturated input. 
In this configuration, we can exploit the simplicity of $g(\cdot)$ by designating edge values $r[k]=\pm\Lambda$ to represent all ADC inputs outside the dynamic range and, thus, do not require knowledge about $r_a[k]$ to evaluate gradients
in the backwards pass. Slight approximation occurs when $\alpha\,r_a[k]$ resides very closely below the saturation threshold, which, however, is negligible with sufficiently fine quantization.

During backpropagation, this approach hence disregards any samples identified to fall outside the dynamic range, as they do not contribute to the overall gradient. This may slow down the neural network learning, specifically, in the beginning, the stronger the saturation by the SI input.    

With partial ADC modeling using $g(\cdot)$, a linear approximation of the ADC's step function takes place within the dynamic range. Strictly speaking, this resembles an approximation according to the straight-through-estimation (STE) principle \cite{yin2019understanding} in the field of quantized neural networks. STE approximates a quantized activation function with a continuous proxy function to compute meaningful gradients during backpropagation. 

\subsection{Straight-Through Estimation (STE)}
\label{sec:ste}
\begin{figure}[h]
    \centering
    \psfrag{Neural Net}{Neural Net}
    \psfrag{ADC}{ADC}
    \psfrag{ADC Noise}{ADC Noise}
    \psfrag{LNA}{LNA}
    \psfrag{SIC}{SIC}
    \psfrag{y[k]}[c][l]{$y_a[k]$}
    \psfrag{ar[k]}{$\alpha r_a[k]$}
    \psfrag{s[k]}{$s[k]$}
    \psfrag{r[k]}{$r[k]$}
    \psfrag{dr[k]}{$\Delta r[k]$}
    \includegraphics{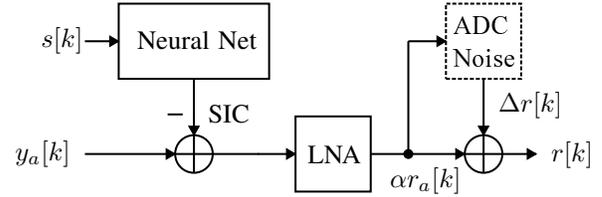}
    \caption{Setup for simulating STE training.}
    \vspace{-1ex}
    \label{fig:onlineSTEsim}
\end{figure}

The aforementioned STE proxy is frequently chosen as a simple neutral linear function. Therefore, we may also extend the linear STE approximation of just the dynamic range of the ADC to additionally include the saturation range.
As shown by Fig. \ref{fig:onlineSTEsim}, this approach translates to omitting the presence of the ADC entirely during backpropagation, which naturally does not entail any additional implementation effort.

This simplicity, however, comes at the cost of potentially large discrepancy between the actual ADC layer nonlinearity in the forward pass and the simplified proxy in the backward pass, specifically, in the undermodeled saturation range. Yet, stable convergence of SIC model learning can be foreseen, since the saturated residual  $r[k]$ of the forward pass is only ever equal to or smaller than the $\alpha r_a[k]$ input to the ADC and, hence, this type of STE generally yields gradients smaller but never larger than without approximation. We therefore expect this STE approach to cope well with initial saturation of the ADC and to not critically impede the desired SIC learning. Moreover, with optimization progress and successful SIC, the STE approximation in the saturation range will diminish.

\subsection{Automatic Gain Control (AGC) based strategy}
\begin{figure}[h]
    \centering
    \psfrag{Neural Net}{Neural Net}
    \psfrag{ADC Noise}{ADC Noise}
    \psfrag{AGC}{AGC}
    \psfrag{ADC}{ADC}
    \psfrag{SIC}{SIC}
    \psfrag{al[k]}{\!\!$1/\alpha[k]$}
    \psfrag{agra}{$g(\alpha r_a[k])$}
    \psfrag{y[k]}{$y_a[k]$}
    \psfrag{ar[k]}{$\alpha r_a[k]$}
    \psfrag{s[k]}{$s[k]$}
    \psfrag{r[k]}{$\tilde{r}[k]$}
    \psfrag{drt[k]}{$\Delta \Tilde{r} [k]$}
    \psfrag{dr[k]}{$\Delta \tilde{r}[k]$}
    \psfrag{rt[k]}{$r[k]$}
    \psfrag{ara[k]}{\!$\alpha[k]r_a[k]$}
    \includegraphics{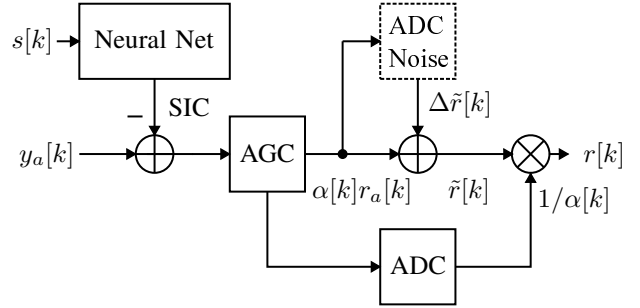}
    \caption{Setup for simulating training with AGC.}
    \label{fig:AGC}
\end{figure}

If a transceiver is equipped with AGC in place of the plain LNA, as shown by Fig.~\ref{fig:AGC},
that alone might be enough to avoid initial saturation. Moreover, if the AGC continuously holds an appropriate signal level throughout optimization, even as the residual $r_a[k]$ becomes very small via SIC, quantization errors are mitigated, potentially allowing for better performance compared to a system without AGC. 

When operating an AGC during model learning, however, backpropagation needs to be informed of the time variable gain $\alpha[k]$ at all times in order to employ a gain-compensated $r[k] = \tilde{r}[k]/\alpha[k]$ instead of the observed residual $\tilde{r}[k]$, such that the gain-compensated residual $r[k]$ suitably diminishes over the course of learning.
While AGC systems commonly provide this gain information, usually referred to as received signal strength indicator (RSSI) $\alpha[k]$, an additional ADC is required to observe the quantity for further use on the digital side. 
For realistic simulation, a similar technique as used for the ''ADC Noise'' block is thus used for AGC gain quantization, however, not precisely shown in Fig.~\ref{fig:AGC} for the sake of simplicity.

\subsection{Digital-Training Approach (DTA)}
As another baseline configuration for avoiding ADC saturation during SIC learning, both LNA and AGC circuitry also can be omitted \cite{kiayani2016active} or the transmission power can be reduced during SI acquisition (with potential change of PA nonlinearity as compared to actual transmission) \cite{kwak2019comparative}.
For reference, we here adopt the strategy of using recordings $y[k]$ of $y_a[k]$ without LNA gain to train a model for digital SIC. In a second step the learned model is employed in the digitally-assisted SIC scheme of Fig.~\ref{fig:modelBlockDiagram}. The core idea of this procedure is that a model trained on the digital $y[k]$ may still manage to predict the analog $y_a[k]$ reasonably well, perhaps better than the $y[k]$, since there is limited possibility for unintentional modeling of the fine-grained nonlinear distortion of the quantization step-function with our small model capacity. 

\begin{figure}[h]
    \centering
    \psfrag{Model}{Model}
    \psfrag{ADC}{ADC}
    \psfrag{SIC}{SIC}
    \psfrag{s[k]}{$s[k]$}
    \psfrag{y(t)}{$y_a[k]$}
    \psfrag{y[k]}{$y[k]$}
    \psfrag{r[k]}{$r[k]$}
    \includegraphics{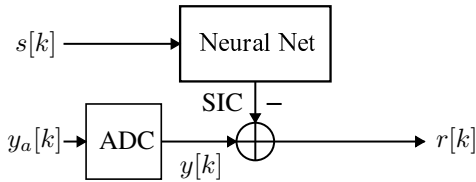}
    \caption{Setup for digital SIC training.}
    \vspace{-0.5ex}
    \label{fig:trainingBased}
\end{figure}

Fig. \ref{fig:trainingBased} depicts the training arrangement, where, different to before, SIC is executed in digital domain. As a consequence, utilization of $r[k]$ during backpropagation is now relieved of any concerns regarding ADC induced distortions. Note that during the recording of training data $y[k]$, the analog SIC in Fig. \ref{fig:modelBlockDiagram} must be suspended. This also implies that training has to be carried out offline, which may be undesirable.

%% file: 5_eval.tex
\section{Experimental Evaluation}
\label{sec:eval}

All system configurations are optimized for $6000$ epochs using a training corpus generated as described in \cite{enzner2024neural}, consisting of $10$ signal sequences of length $4096$ each, where the transmit signal $s[k]$ is normalized to unit variance. The size of the training corpus is deemed appropriate with respect to the relatively small model capacity. We choose the Adam~\cite{Kingma2014AdamAM} optimizer with MSE loss and learning rate $0.03$. The ADC is simulated with $B=12$ bits resolution and $\Lambda = 1$ saturation.

\subsection{Learning Behavior and Performance of SIC}
\label{sec:evalSIC}

The SOI is absent during SIC learning, i.e., $x(t)=0$ in Fig.~\ref{fig:modelBlockDiagram}. For BPAD and STE, the LNA will be considered with different gains \{$30, 40, 50$\}\,dB to demonstrate scenarios with different severity of ADC saturation. We employ a frame-wise implementation of the AGC system with frame length $N_{\text{AGC}}=64$, a square-law detector~\cite{whitlow2003design} and a set-point determined experimentally to provide an appropriate signal level that mostly avoids saturation. The range of available AGC gains is configured to $[\text{-}20, 40]$\,dB to enable proper attenuation of initial SI, while providing for sufficient gain in the later phase of SIC optimization.
The ADC used to observe $\alpha[k]$ is configured identically to the primary ADC of the receive path, but with a dynamic range that suits the gain range.  For DTA, the LNA is bypassed for avoiding the ADC saturation while recording the training data.

Fig.~\ref{fig:evalMSE} (a) firstly compares the development of the power of the SIC residual $r_a[k]$ over the course of neural network learning, where BPAD and STE were configured with LNA gain of $30$\,dB. The dashed line at -$35$\,dBm indicates the approximate signal power where ADC saturation is just avoided. In reality, as discussed before, $r_a[k]$ is not observable during learning, but the simulation allows its inspection. 
We observe similar performance of all learning strategies. In all cases, the SIC residual starts around -$15$\,dBm (i.e., the SI level) and falls to around -$63$\,dBm, corresponding to nearly $50$\,dB digital SI attenuation.
Notably, there seems to be no significant advantage of AGC efforts in this SIC analysis.

To better illustrate different sensitivities of BPAD and STE, Fig.~\ref{fig:evalMSE} (b) further compares the power of $r_a[k]$ over the course of learning for LNA gains $\{30, 40, 50\}$\,dB and thus additionally more pronounced effects of (initial) ADC saturation. 
With LNA=$40$dB, BPAD firstly indicates a somewhat decelerated learning process and it fails to converge with LNA gain of $50$dB, where a critical number of $r_a[k]$ samples fall outside the modeled dynamic range. Meanwhile, STE is seemingly unimpressed by even the strongest saturation at LNA=$50$dB, i.e., exhibiting only minor differences across all three gains.

\subsection{Demodulation Performance in Terms of Bit-Error Rate}

\begin{figure}[t]
    \centering
    (a) LNA gain $30$\,dB\\
    \includegraphics[width=8.8cm, height=4.50cm]{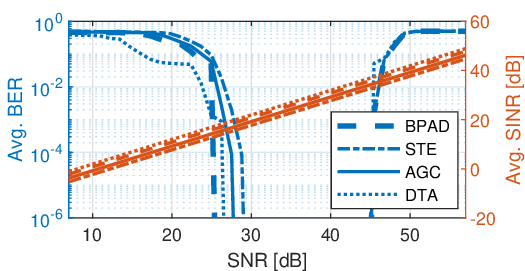}\\[1ex]
    (b) LNA gain $40$\,dB\\
    \includegraphics[width=8.8cm, height=4.50cm]{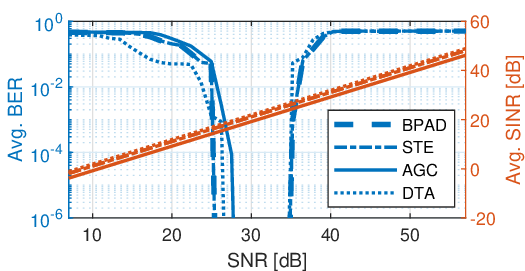}\\
    \caption{Demodulation performance for various LNA gains.}
    \vspace{-2ex}
    \label{fig:evalBER}
\end{figure}

In order to evaluate the BER, $r_a[k]$ signals, obtained with learned SIC models and applied to signals from a separate test dataset, are mixed with SOI $x_a[k]$ at different signal-to-noise ratios SNR=$P_{x,a}/P_{n,a}$. The mixed signals are then demodulated using Matlab's WLAN Toolbox\footnote{Demodulation functions from https://de.mathworks.com/help/wlan/ug/802-11n-packet-error-rate-simulation-for-2x2-tgn-channel.html} to recover the original bit sequences.
Where packet detection fails, we assign BER\,=$0.5$.
Performance in terms of an averaged BER versus SNR is shown in Fig.~\ref{fig:evalBER}. The underlying signal-to-inference-and-noise ratios SINR=$P_{x,a}/P_{r,a}$ after SIC is depicted on the right y-axis. It refers to the SOI power w.r.t.\ residual SI and noise disturbances before demodulation.

\begin{figure}[t]
    \centering
    (a) LNA gain $30$\,dB\\
    \psfrag{Residual Power [dBm]}{$r_a[k]$ power [dBm]}
    \psfrag{Epochs}{Epochs}
    \includegraphics[width=8.8cm, height=4.50cm]{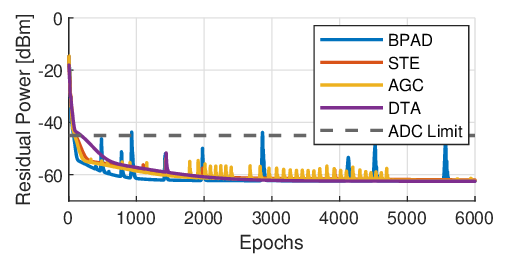}\\[1ex]
    (b) LNA gains \{$30, 40, 50$\}\,dB\\
    \includegraphics[width=8.8cm, height=4.50cm]{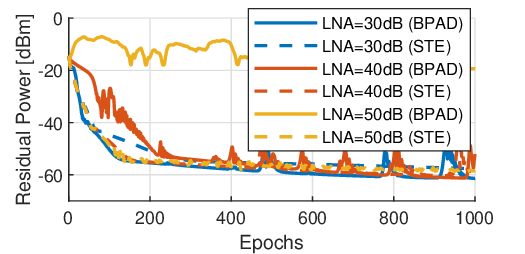}\\
    \caption{Learning behavior of SIC for various LNA gains.}
    \vspace{-2ex}
    \label{fig:evalMSE}
\end{figure}

Fig.~\ref{fig:evalBER}~(a) compares the performance of all system options, where LNA gain of $30$\,dB is used for BPAD and STE. In accordance with results from Fig.~\ref{fig:evalMSE} (a), the BER performance turns out to be similar across the different systems. BER values start to fall around SNR\,=\,$20$\,dB based on the underlying SINR~$\geq10$\,dB. The desired BER\,$\simeq 0$ is achieved with all systems in the operating range of SNR\,=\,$[30, 45]$\,dB. Performances of BPAD, STE and DTA systems then quickly degrade again for SNR\,$>45$\,dB due to SOI saturation by the ADC,
while the AGC system continuous to maintain low BER by adjusting the signal level properly. Note that this advantage is now provided by the AGC mainly operating on the SOI, i.e., not on the SIC as in the former model learning process.

Fig.~\ref{fig:evalBER}~(b) finally summarizes end-to-end BER performance for a larger $40$\,dB LNA gain, where system parameters for AGC and DTA are identical to Fig.~\ref{fig:evalBER}~(a).
We observe marginally improved but fundamentally similar performance for BPAD and STE up until SNR=$30$\,dB. Most prominently, the larger LNA gain causes earlier SOI saturation by the ADC, thus narrowing the operational SNR range to a smaller window merely around $30$\,dB (except for the AGC system).

%% file: 6_concl.tex
\section{Conclusions}
\label{sec:concl}
The learning of a digital system function for analog (i.e., digitally-assisted) SIC rests upon the residual in the digital domain and is complicated by the ADC function in between. This study presented a range of viable system options regarding the modeling of the ADC with different amenities and limitations: an intuitive yet fragile backpropagation-through-ADC (BPAD) approach, a surprisingly simple and robust straight-through-estimation (STE) technique, a more costly AGC-based system, and an offline digital-training-based approach (DTA). For~low system complexity and online adaptivity of SIC systems, the STE approach turns out favorable in our simulation.

%% file: 2025_IBFD_ADC.bbl
\begin{thebibliography}{10}
\providecommand{\url}[1]{#1}
\csname url@samestyle\endcsname
\providecommand{\newblock}{\relax}
\providecommand{\bibinfo}[2]{#2}
\providecommand{\BIBentrySTDinterwordspacing}{\spaceskip=0pt\relax}
\providecommand{\BIBentryALTinterwordstretchfactor}{4}
\providecommand{\BIBentryALTinterwordspacing}{\spaceskip=\fontdimen2\font plus
\BIBentryALTinterwordstretchfactor\fontdimen3\font minus \fontdimen4\font\relax}
\providecommand{\BIBforeignlanguage}[2]{{%
\expandafter\ifx\csname l@#1\endcsname\relax
\typeout{** WARNING: IEEEtran.bst: No hyphenation pattern has been}%
\typeout{** loaded for the language `#1'. Using the pattern for}%
\typeout{** the default language instead.}%
\else
\language=\csname l@#1\endcsname
\fi
#2}}
\providecommand{\BIBdecl}{\relax}
\BIBdecl

\bibitem{Heino_2015}
M.~Heino \emph{et~al.}, ``Recent advances in antenna design and interference cancellation algorithms for in-band full duplex relays,'' \emph{IEEE Comm. Mag.}, vol.~53, no.~5, pp. 91--101, 2015.

\bibitem{Herd_2019}
K.~Kolodziej, B.~Perry, and J.~Herd, ``In-band full-duplex technology: Techniques and systems survey,'' \emph{IEEE Trans. Microwave Theory Techn.}, vol.~67, no.~7, pp. 3025--3041, 2019.

\bibitem{Smida2023}
B.~Smida, A.~Sabharwal, G.~Fodor, G.~Alexandropoulos, H.~Suraweera, and C.~Chae, ``Full-duplex wireless for {6G}: Progress brings new opportunities and challenges,'' \emph{IEEE Jrnl. Selected Areas Commu.}, vol.~41, no.~9, pp. 2729--2750, 2023.

\bibitem{korpi2014full}
D.~Korpi, T.~Riihonen, V.~Syrj{\"a}l{\"a}, L.~Anttila, M.~Valkama, and R.~Wichman, ``Full-duplex transceiver system calculations: Analysis of {ADC} and linearity challenges,'' \emph{IEEE Transactions on Wireless Communications}, vol.~13, no.~7, pp. 3821--3836, 2014.

\bibitem{kiayani2016active}
A.~Kiayani, L.~Anttila, and M.~Valkama, ``Active {RF} cancellation of nonlinear {TX} leakage in {FDD} transceivers,'' in \emph{IEEE Global Conf. Signal and Information Processing (GlobalSIP)}, 2016, pp. 689--693.

\bibitem{Valkama_2018}
A.~Kiayani \emph{et~al.}, ``Adaptive nonlinear {RF} cancellation for improved isolation in simultaneous transmit–receive systems,'' \emph{IEEE Trans. Microw. Theory Techn.}, vol.~66, no.~5, pp. 2299--2312, 2018.

\bibitem{liu2017digitally}
Y.~Liu, X.~Quan, W.~Pan, and Y.~Tang, ``Digitally assisted analog interference cancellation for in-band full-duplex radios,'' \emph{IEEE Communications Letters}, vol.~21, no.~5, pp. 1079--1082, 2017.

\bibitem{kwak2019comparative}
J.~W. Kwak, M.~S. Sim, I.-W. Kang, J.~S. Park, J.~Park, and C.-B. Chae, ``A comparative study of analog/digital self-interference cancellation for full duplex radios,'' in \emph{IEEE Asilomar Conf. on Signals, Systems, and Computers}, 2019, pp. 1114--1119.

\bibitem{riihonen2012analog}
T.~Riihonen and R.~Wichman, ``Analog and digital self-interference cancellation in full-duplex {MIMO-OFDM} transceivers with limited resolution in {A/D} conversion,'' in \emph{IEEE Asilomar Conf. on Signals, Systems and Computers}, 2012, pp. 45--49.

\bibitem{xing2020comprehensive}
J.~Xing, S.~Ge, Y.~Liu, Z.~Cui, and J.~Meng, ``Comprehensive analysis of quantization effects on digital-controlled adaptive self-interference cancellation system,'' \emph{IEEE Access}, vol.~8, pp. 75\,772--75\,784, 2020.

\bibitem{enzner2024neural}
G.~Enzner, A.~Chinaev, S.~Voit, and A.~Sezgin, ``On neural-network representation of wireless self-interference for inband full-duplex communications,'' \emph{arXiv preprint arXiv:2410.00894}, 2024.

\bibitem{Stimming2019}
A.~Kristensen, A.~Burg, and A.~Balatsoukas-Stimming, ``Advanced machine learning techniques for self-interference cancellation in full-duplex radios,'' in \emph{Asilomar Conf. Signals, Systems, and Computers}, 2019, pp. 1149--1153.

\bibitem{bengio2013estimating}
Y.~Bengio, N.~L{\'e}onard, and A.~Courville, ``Estimating or propagating gradients through stochastic neurons for conditional computation,'' \emph{arXiv preprint arXiv:1308.3432}, 2013.

\bibitem{yin2019understanding}
P.~Yin, J.~Lyu, S.~Zhang, S.~Osher, Y.~Qi, and J.~Xin, ``Understanding straight-through estimator in training activation quantized neural nets,'' \emph{arXiv preprint arXiv:1903.05662}, 2019.

\bibitem{Qureshi_2023}
I.~Qureshi and S.~Asghar, ``A systematic review of the {IEEE-802.11} standard’s enhancements and limitations,'' \emph{Wireless Personal Comm.}, vol. 131, no.~4, pp. 2539--2572, 2023.

\bibitem{abadi2016tensorflow}
M.~Abadi, P.~Barham, J.~Chen, Z.~Chen, A.~Davis, J.~Dean, M.~Devin, S.~Ghemawat, G.~Irving, M.~Isard \emph{et~al.}, ``Tensorflow: a system for large-scale machine learning,'' in \emph{USENIX Symposium on Operating Systems Design and Implementation}, 2016, pp. 265--283.

\bibitem{Kingma2014AdamAM}
D.~P. Kingma and J.~Ba, ``Adam: A method for stochastic optimization,'' \emph{Intl.\ Conf.\ Learning Representations}, vol. abs/1412.6980, 2014.

\bibitem{whitlow2003design}
D.~Whitlow, ``Design and operation of automatic gain control loops for receivers in modern communications systems,'' \emph{Microwave Journal}, vol.~46, no.~5, pp. 254--256, 2003.

\end{thebibliography}
